# General Predictive Framework for Droplet Detachment Force

Muhammad Subkhi Sadullah[#], Yinfeng Xu[#], Sankara Arunachalam, Himanshu Mishra*

Environmental Science and Engineering (EnSE) Program, Biological and Environmental Science and Engineering (BESE) Division, King Abdullah University of Science and Technology (KAUST), Thuwal 23955-6900, Kingdom of Saudi Arabia

Water Desalination and Reuse Center (WDRC), KAUST

Center for Desert Agriculture (CDA), KAUST

[#]These authors contributes equally
*Corresponding author: himanshu.mishra@kaust.edu.sa

**Abstract:**

Liquid droplets hanging from solid surfaces are commonplace, but their physics is complex. Examples include dew or raindrops hanging onto wires or droplets accumulating onto a cover placed over warm food or windshields. In these scenarios, determining the force of detachment is crucial to rationally design technologies. Despite much research, a quantitative theoretical framework for detachment force remains elusive. In response, we interrogated the elemental droplet–surface system via comprehensive laboratory and computational experiments. The results reveal that the Young–Laplace equation can be utilized to accurately predict the droplet detachment force. When challenged against experiments with liquids of varying properties and droplet sizes, detaching from smooth and microtextured surfaces of wetting and non-wetting chemical make-ups, the predictions were in an excellent quantitative agreement. This study advances the current understanding of droplet physics and will contribute to the rational development of technologies.

# I. Introduction

Droplet attachment to and detachment from solid surfaces is ubiquitous in nature, e.g., morning dew drops on spider webs [1], grass blades [2], and the compound eyes of insects [3] and sea spray on the exoskeletons of marine skaters [4] and desert beetles [5]. From an industrial perspective, droplet–surface interactions are important in the application of foliar pesticides/nutrients [6], separation processes [7], heat transfer [8], interfacial chemistry [9], atmospheric water harvesting [1,10], and the cleaning of windowpanes, windshields, and solar cells [11]. Some of these applications require the droplets to stick to a surface (e.g., foliar sprays [6,12]), while others require their prompt removal (e.g., windshields [11]). Therefore, it is crucial to understand the mechanisms associated with the droplet detachment force. In this respect, it has been reported that the magnitudes of normal and lateral droplet detachment forces are dissimilar; for example, sessile droplets of water/oil may slide on a perfluorinated kitchen pan when tilted but may not detach when the pan is held upside down [13,14,15].

Several experimental measurements have been employed to describe the lateral detachment of droplets, including the advancing ($\theta_A$) and receding ($\theta_R$) contact angle [16], the roll-off angle [17,18], and the direct lateral movement of droplets, which are generally compared using the Furmidge relation $F_{\text{lat}} = L_p \times \gamma(\cos\theta_R - \cos\theta_A)$, where $L_p$ and $\gamma$ are the droplet base perimeter and the liquid surface tension, respectively[17,19]. For the normal detachment force, ring tensiometry has been employed at the millimeter length scale [20,21,22], while scanning droplet adhesion microscopy [23] and droplet force apparatus techniques [24] have more recently been used to produce data at a micrometer resolution with a nanonewton sensitivity. Ferrofluids have also been utilized to measure the droplet detachment force under a magnetic field [25,26]. Despite this body of research, quantitative insights into the mechanisms associated with droplet detachment are lacking [15,25,27,28,29,30]. Complications can arise when seeking to interpret experimental data because the detachment force is found to be sensitive to the size and volume of the droplet [31,32] or to the residue left behind by the detaching droplet. In the latter case, detachment initiates at the solid–liquid–vapor interface but proceeds and breakage in a manner similar to Tate's law and/or a dripping faucet [32,33]. Therefore, it is important to account for the contributions of the many subtle factors that influence the droplet detachment force.

The Young–Dupré equation, introduced over 153 years ago, has been utilized to theoretically describe droplet–solid surface adhesion [32,33]. It estimates the work of adhesion using the formula $W_{\text{SL}} = \gamma(1 + \cos\theta_e)$, where $\theta_e$ is the droplet contact angle at thermodynamic equilibrium [32]. This approach assumes an idealized scenario in which the liquid droplet completely detaches from the solid surface without any deformation. Some researchers have argued the use of the $W_{\text{SL}}$ expression with the receding angle $\theta_R$ has a significantly stronger correlation with the measured detachment force [14, 20, 34]. Therefore, the detachment force $F_D$ can be expressed following Tate's law as

$$F_D = 2\pi R_i \gamma (\cos \theta_R + 1), \tag{1}$$

where $R_i$ is the base radius of the initial droplet. Assuming that the droplet is hemispherical, $R_i$ can be expressed as

$$R_i = \left[\frac{3V}{\pi} \frac{\sin \theta_R (1+\cos \theta_R)}{(1-\cos \theta_R)(\cos \theta_R + 2)}\right]^{1/3}. \tag{2}$$

In their pioneering study, Tadmor and co-workers developed the centrifugal adhesion balance (CAB) technique, which allows for the tilt-free rotation of a pendant droplet so that the centrifugal force increases its weight (via the applied "body force") and records the droplet shape when it detaches [35]. However, in the CAB approach, it is not entirely clear whether the advancing or receding contact angle should be used. Tadmor and his research team also contended that $R_i$ in Eq. (1) should be replaced by the droplet radius at the moment the critical body force is reached ($R_D$), after which the droplet diameter starts shrinking spontaneously before detachment [35]. Due to the significance of $\theta_R$ in terms of droplet detachment, we can modify Eq. (1) by replacing $R_i$ with $R_D$:

$$F_D = 2\pi R_D \gamma (\cos \theta_R + 1). \tag{3}$$

Here, it should be noted that $R_D$ must be measured experimentally; therefore, Eq. (3) cannot be used to predict the detachment force *a priori*. Some researchers have empirically modified the Young–Dupre equation to derive a "universal fit" for $F_D$ [25]. Following a different approach, Butt and co-workers have argued that the detachment force of a droplet sitting on a smooth surface pulled by an actuator is affected by the surface tension and Laplace pressure forces at the moment before detachment [27]:

$$F_D = 2\pi R_D \gamma \sin \theta_R - \pi R_D^2 \Delta P, \tag{4}$$

where $\Delta P = \gamma \left(\frac{1}{R_1} + \frac{1}{R_2}\right)$ and $R_1$ and $R_2$ are the radii of curvatures at the interface. Eq. (4) can be used to calculate the force experienced by the droplet when $R_D$ and $\Delta P$ are known. However, because $R_D$ and $\Delta P$ need to be measured by analyzing the droplet shape before detachment, Eq. (4) cannot be used to predict $F_D$.

Given the proliferation of experimental techniques, competing models, and scientific debate, an encompassing theoretical framework for droplet detachment is needed to draw together the expansive body of experimental work. To this end, we combine laboratory experiments, theory, and computation in this paper to answer the following elemental questions:

1. Can detachment force $F_D$ be experimentally induced by exerting a normal body force on (or increasing the weight of) a pendant droplet placed on a wetting (or non-wetting) surface with a smooth (or microtextured) topography?
2. Does $F_D$ depend only on the work of adhesion ($W_{SL}$) prescribed by Eq. (1), or does it also depend on the stability/curvature of the liquid–vapor interface during detachment?
3. Is it possible to simulate laboratory experiments in silico and develop an encompassing theoretical framework to accurately predict $F_D$?
4. Can this framework predict $F_D$ for different detachment causes, such as volume increase or reduction of interfacial tension (e.g., oil–water–solid system)?

Our study reveals that the droplet detachment force $F_D$ is not a function of the work of adhesion as prescribed by the Young–Dupré equation (Eq. 1). Instead, $F_D$ is related to instability at the liquid-vapor interface, which can be predicted by solving the Young–Laplace equation (YLE). This theoretical framework quantitatively captures the $F_D$ of pendant droplets for multiple probe liquids detaching from flat and microtextured surfaces with varying chemical make-ups; it also affords encompassing insights into droplet detachment in scenarios where gravity and/or buoyancy are relevant.

## II. Results

**Samples and probe liquids:** To study $F_D$, we employed smooth and textured substrates with chemical make-ups ranging from wetting to non-wetting. The substrates included silanized $SiO_2/Si$ wafers, flat polystyrene, and microtextured $SiO_2/Si$ (Fig. 1). The silanized $SiO_2/Si$ samples were prepared by functionalizing $SiO_2/Si$ wafers with (3-aminopropy)triethoxysilane, trichloro(octadecyl)silane, 11-bromoudecyltrichlorosilane, 10-undecenyltrichlorosilane, and 10-undecenyltrichlorosilane, following a recently reported method [36]. The polystyrene samples were obtained commercially and used without any surface modification. For the microtextured surfaces, the photolithography and dry etching of $SiO_2$ and Si layers were utilized to create arrays of cylindrical pillars with a diameter, height, and center-to-center distance of 20 μm, 50 μm, and 25 μm, respectively. After microfabrication, the surface was functionalized with perfluorodecyltrichlorosilane (FDTS) to render it hydrophobic, following a recently reported process [37]. These chemical and physical treatments were employed to modify the surface wettability to produce a wide range of apparent contact angles for a more comprehensive analysis of droplet adhesion forces. The microtextured surfaces were designed specifically not to exhibit superhydrophobicity to ensure that the analysis of pendant droplet detachment was possible.

The samples were stored in glass petri dishes in a $N_2$ flow cabinet and, before their use, they were rinsed with ethanol and water and dried with $N_2$ gas. Water and ethylene glycol were used as the probe liquids. To characterize the wettability of the samples, we measured the apparent advancing ($\theta_A$) and receding ($\theta_R$) angles of water and ethylene glycol on the samples using a goniometer. The measurement of the apparent contact angles involved placing a drop (2–6 µl) on the surface and then recording the advancing and receding angles with the addition and then removal of 10 µl to the drop at a rate of 0.2 µL/s.

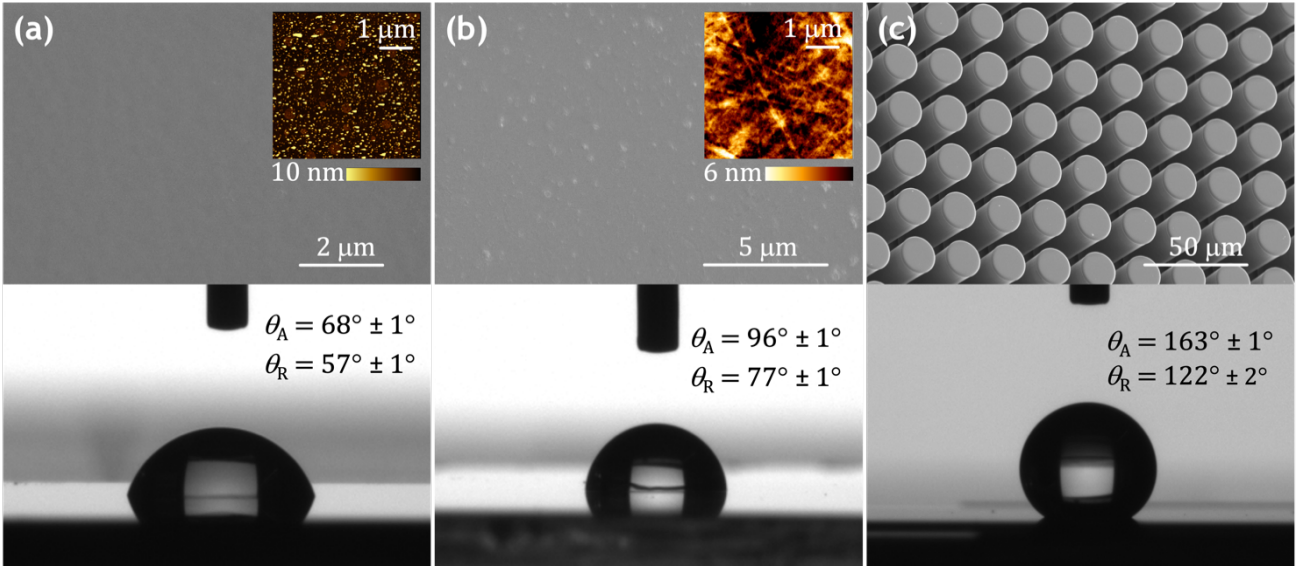

**Fig. 1. Surface characterization.** Samples of various wettabilities were tested in this study, including (a) a silanized Si wafer, (b) flat polystyrene, and (c) an SiO$_2$ micropillar array. The top and bottom panels show the scanning electron micrographs and the measured contact angles of the samples. The insets in panels (a) and (b) present the atomic force micrographs where the color contrasts represent the topographical variation of the surface.

**Laboratory experiments on normal droplet detachment:** The normal detachment force $F_D$ was directly measured using the CAB technique. This entails the application of a centrifugal force to manipulate the droplet body force (or increase its weight) and detach it. CAB ensures that the droplet does not tilt during normal detachment, as illustrated in Fig. 2(a) [35]. This is because the sum of the lateral forces is zero ($g \sin \alpha = \omega^2 R \cos \alpha$; see the free-body diagram in the inset of Fig. 2a). As a result, the effective normal acceleration is given by $g_{\text{eff}} = g / \cos \alpha$, where $g$ and $\alpha$ are the gravitational acceleration and the angle between the sample and the horizontal line. The normal force is then given by $F_g = mg_{\text{eff}}$, and we expressed it in a non-dimensional form by normalizing it with $\gamma \sqrt[3]{V}$ as follows:

$$\tilde{F}_g = \frac{mg_{\text{eff}}}{\gamma \sqrt[3]{V}} = \frac{\rho g_{\text{eff}} V^{2/3}}{\gamma}, \tag{5}$$

where *m*, *V,* and *ρ* are the droplet mass, volume, and density, respectively. In a typical CAB experiment, the body force is increased incrementally until detachment occurs. Snapshots of the droplet detachment experiments conducted using the CAB approach are available in SI Movie and Fig. 2(b). The corresponding force is then registered as detachment force $F_D$, which can be written in a non-dimensional form as follows:

$$\tilde{F}_D = \frac{\rho g_c V^{2/3}}{\gamma}, \tag{6}$$

where $g_c$ is the critical acceleration. This non-dimensional form was employed in the present study to disentangle the dependence of $F_D$ on the droplet volume, the surface tension of the probe liquid, the surface wettability, and the microtexture. To this end, we first investigated whether the non-dimensional detachment force was independent of the droplet volume. By varying the water droplet volume from $V \approx 3$–$13$ μl on substrates of different wettability, we established that $\tilde{F}_D$ is constant for a given $\theta_R$ (Fig. 2c). We then computationally and experimentally studied the detachment process as a function of $\theta_R$.

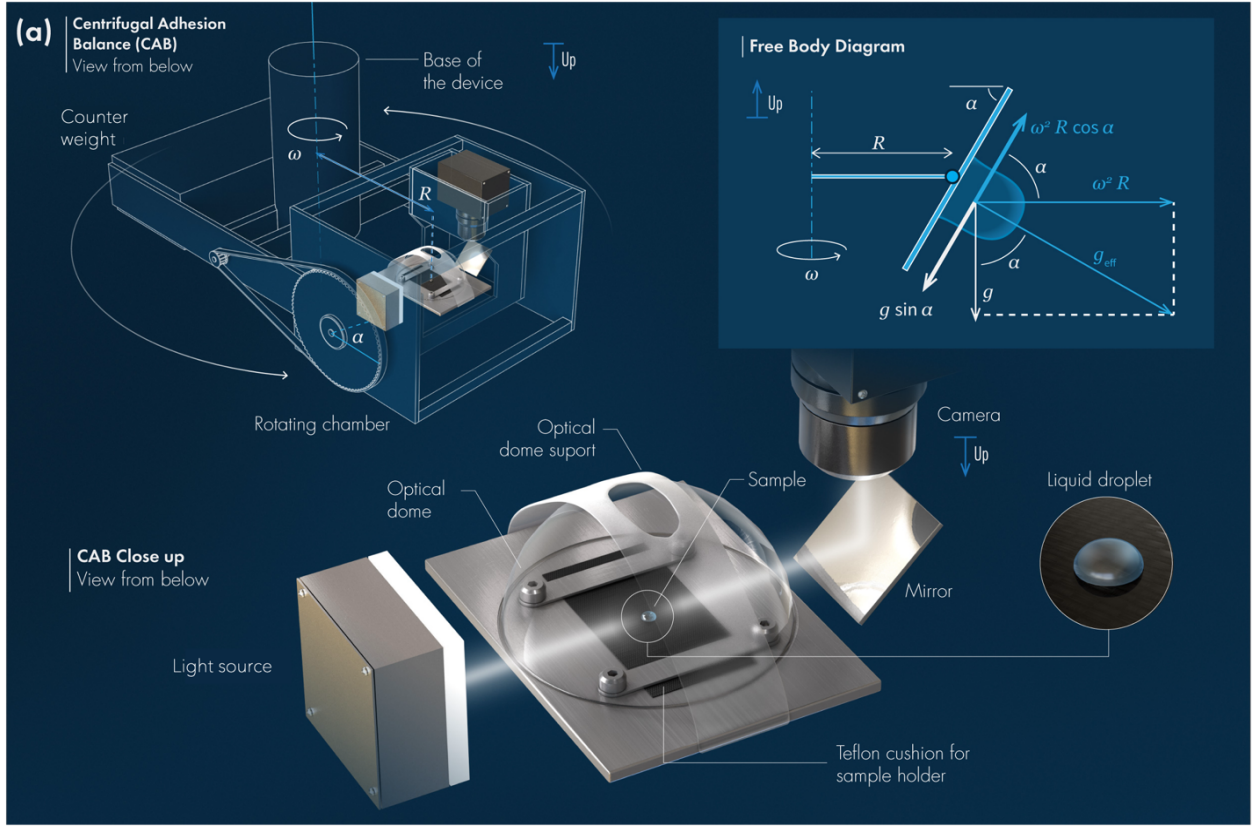

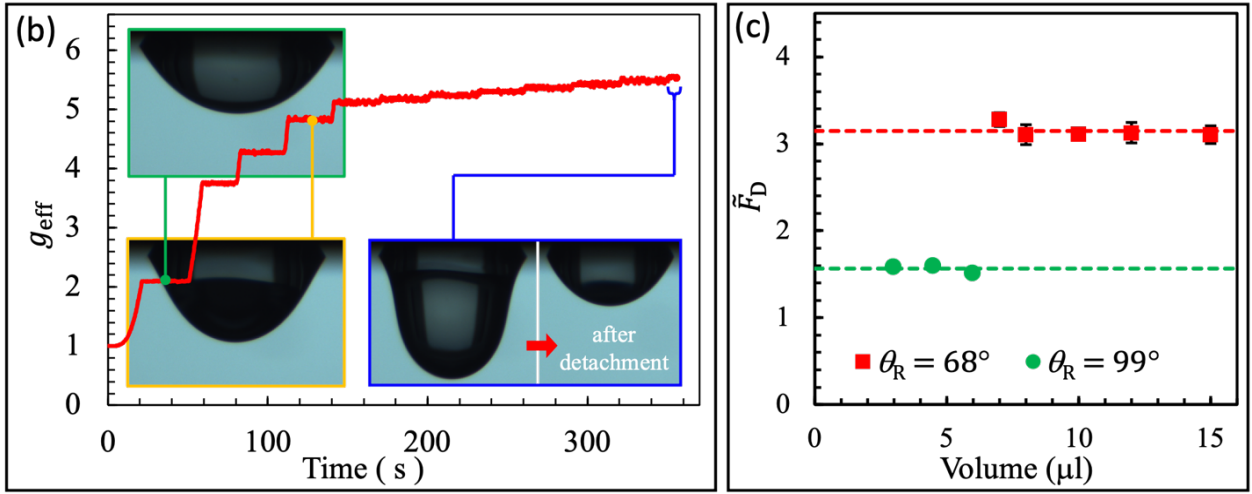

**Fig. 2. Centrifugal Adhesion Balance (CAB) experiments.** (a) Schematic illustration of the CAB system. The inset shows a free-body diagram of a pendant droplet. Zero lateral force is established by ensuring that $g \sin \alpha = \omega^2 R \cos \alpha$. (b) In a typical measurement set-up, the body force of a pendant drop is gradually increased until it detaches and critical acceleration $g_c$ is recorded. Representative snapshots at various times are presented. (c) Non-dimensional detachment force $\tilde{F}_D$ for different receding angles $\theta_R$. The data reveal that, in our volumetric range, $\tilde{F}_D$ does not vary significantly. Note: each point is the average of three $\tilde{F}_D$ measurements of a given volume while the error bars represent their standard deviation. The dashed lines indicate the average $\tilde{F}_D$ value for different volumes.

**Lattice Boltzmann simulations:** To complement our laboratory investigation of droplet detachment, lattice Boltzmann (LB) simulations were performed. The LB algorithm numerically solves the Navier–Stokes and continuity equations to recover the hydrodynamics of a fluid system [38]. These simulations enabled us to capture the droplet detachment from smooth and microtextured surfaces over a wide range of apparent contact angles, which would be difficult and laborious to study experimentally. In a typical *in silico* experiment, $g_{\text{eff}}$ was increased incrementally in a manner similar to the CAB experiment until the droplet detached. This yielded the critical acceleration $g_c$, which could be used to calculate $\tilde{F}_D$ via Eq. (6). Fig. 3 presents a representative comparison of the laboratory experiments with the LB simulations of a pendant drop detaching from a wetting surface. As $\tilde{F}_g$ increased and more liquid volume was drawn downward (compare Fig. 3a–d), the droplet shapes obtained from the simulation accurately reflected those from the experiment. Details of the simulation method are provided in the Methods section and in the Supplementary Information.

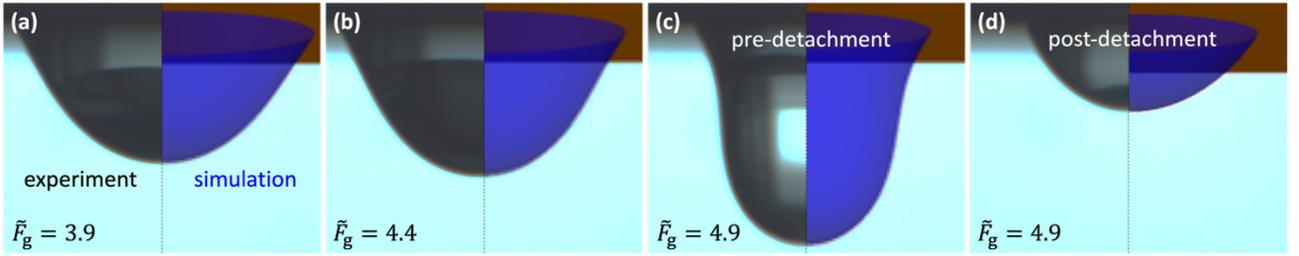

**Fig. 3. A representative juxtaposition of droplet shapes from CAB experiment and LB simulation.** Comparison of droplet shapes obtained from the CAB experiment (left) and LB simulation (right) for different value of body force $\tilde{F}_g$ (a-c). Panels (c) and (d) show the droplet shapes before and after detachment, respectively. The receding angles for both the CAB experiment and LB simulation are similar ($\theta_R \approx 55°$).

**Quantifying the normal droplet detachment force:** Droplet detachment force $\tilde{F}_D$ was quantified based on the experimental and numerical methods using Eq. (6) with $\theta_R$ employed as the pertinent angle. With our experimental and computational approach, we were able to test a wide range of $\theta_R$ (30°–140°) on various surfaces (Table 1). The CAB experiments and subsequent image analysis were used to assess changes in the droplet geometry prior to detachment (SI Movie). The measured detachment force was compared with the predicted detachment force based on Eq. (1) (the green line in Fig. 4) and Eq. (3) (the blue and purple data points). To fairly compare the droplet detachment forces across the various samples (Table 1) and probe liquids of varying volumes, we normalized the forces (Eqs. 1 and 3) with the factor $\gamma\sqrt[3]{V}$. This reflects the fact that, while the theoretical models qualitatively capture the relationship between $\tilde{F}_D$ and $\theta_R$, they fail to describe it quantitatively.

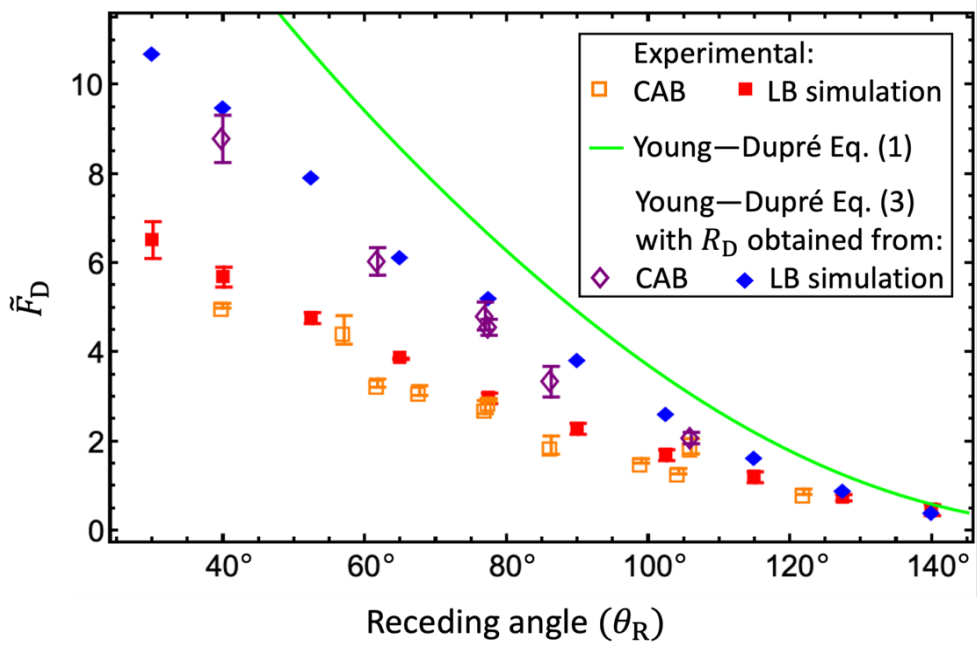

**Fig. 4. The measurement of the detachment force.** Measured $\tilde{F}_D$ calculated using Eq. (6) for different $\theta_R$ obtained from LB simulations (■) and CAB experiments (□). The error bars for the simulation data represent cases where the droplet is still attached or detached, respectively, while they represent the standard deviation of the measurement for the experimental data. The measured values are compared with theoretical predictions using Eq. (1) (plotted as —), and Eq. (3) (plotted as ◆ and ◇). Both Eqs. (1) and (3) are normalized using $\gamma\sqrt[3]{V}$ for this plot.

**Table 1.** A summary of the surface functional groups, probe liquids, and apparent contact angles utilized in the comparative assessment of non-dimensional droplet detachment force $\tilde{F}_D$ (Fig. 4). The surface tension and density were 72.2 mN/m and 997 kg/m³ for water and 35.2 mN/m and 1099 kg/m³ for ethylene glycol, respectively, at 298 K and 1 atm [39]. Note: $\tilde{F}_D$ is obtained using Eq. (6).

| Substrate chemical make-up | Probe liquid | $\theta_A$ | $\theta_R$ | $\tilde{F}_D$ (from CAB) |
|---|---|---|---|---|
| (3-Aminopropy)triethoxysilane on a flat SiO$_2$/Si wafer | Water | 74° ± 1° | 39° ± 1° | 5.0 ± 0.1 |
| Trichloro(octadecyl)silane on a flat SiO$_2$/Si wafer | Ethylene glycol | 69° ± 1° | 57° ± 1° | 4.4 ± 0.2 |
| 11-Bromoudecyltrichlorosilane on a flat SiO$_2$/Si wafer | Water | 101° ± 4° | 62° ± 3° | 3.3 ± 0.1 |
| 11-Bromoudecyltrichlorosilane on a flat SiO$_2$/Si wafer | Water | 97° ± 2° | 68° ± 1° | 3.2 ± 0.1 |
| 10-Undecenyltrichlorosilane on a flat SiO$_2$/Si wafer | Water | 107° ± 2° | 77° ± 1° | 2.8 ± 0.2 |
| Flat polystyrene | Water | 96° ± 1° | 77° ± 1° | 2.9 ± 0.1 |
| Trichloro(octadecyl)silane on a flat SiO$_2$/Si wafer | Water | 126° ± 3° | 86° ± 3° | 1.9 ± 0.2 |
| Trichloro(octadecyl)silane on a flat SiO$_2$/Si wafer | Water | 119° ± 1° | 99° ± 3° | 1.6 ± 0.1 |
| FDTS on a micropillar array on a SiO$_2$/Si wafer | Ethylene glycol | 163° ± 1° | 104° ± 3° | 1.3 ± 0.1 |
| Perfluorodecanethiol on a flat gold-coated SiO$_2$/Si wafer | Water | 117° ± 1° | 106° ± 2° | 1.9 ± 0.2 |
| FDTS on a micropillar array on a SiO$_2$/Si wafer | Water | 163° ± 1° | 122° ± 3° | 0.9 ± 0.1 |

**Theoretical approach:** The laboratory experiments and computer (LB) simulations were in excellent agreement (Fig. 4), with both challenging the common belief that the droplet detachment force is quantitatively related to the work of adhesion. This discrepancy possibly arose because real-world droplet detachment differs considerably from the idealized scenario. In the latter case, droplets with a constant volume ($V_0$), an intrinsic contact angle ($\theta_e$), and spherical sections with curved (liquid–vapor) and flat (liquid–solid) areas ($A_C$ and $A_P$, respectively) obey the geometric relationship $\left(\frac{dA_C}{dA_P}\right)_{V_0} = \cos\theta_e$ [40, 41, 42, 43]. This is only possible when the apparent contact angle does not change significantly from its value at thermodynamic equilibrium. However, as captured in our experimental images (Fig. 2b and 3), droplets undergo dramatic departures from the equilibrium configuration during detachment. Having demonstrated that $\tilde{F}_D$ cannot be quantitatively captured using the classical theory of the work of adhesion and its variations, we present an alternative approach based on the YLE, which relates the pressure difference across a liquid–vapor interface ($\Delta P$) due to its radii of curvature ($R_1$ and $R_2$) and surface tension as [32]

$$\Delta P = \gamma \left(\frac{1}{R_1} + \frac{1}{R_2}\right). \quad (7)$$

For an axisymmetric system, such as our pendant drop, Eq. (7) can be expressed in Cartesian coordinates as

$$\Delta P = \gamma \left(\frac{d^2z/dx^2}{(1+(dz/dx)^2)^{3/2}} + \frac{1}{z(1+(dz/dx)^2)^{1/2}}\right), \quad (8)$$

where $x$ and $z$ are the horizontal and vertical axes, respectively. The solution for $x(z)$ from Eq. (8) yields the droplet shape. In order to solve this non-linear second-order ordinary differential equation (ODE), we follow O'Brien and van den Brule's approach [44] by introducing new variables as a function of the droplet's contour path $s$:

$$\Delta P = \gamma \left( \frac{d\phi}{ds} + \frac{\sin \phi}{x} \right), \tag{9a}$$

$$\frac{dx}{ds} = \cos \phi, \tag{9b}$$

$$\frac{dz}{ds} = \sin \phi, \tag{9c}$$

where $\phi$ is the angle that any arbitrary point on the path makes with the vertical axis. Next, we consider the pressure difference inside across the interface $\Delta P$, which is given by the pressure at the apex and the hydrostatic pressure, meaning that Eq. (9a) can be written as

$$\frac{2\gamma}{R_0} - \rho g_{\text{eff}} z = \gamma \left( \frac{d\phi}{ds} + \frac{\sin \phi}{x} \right), \tag{10}$$

where $R_0$ is the radius of curvature at the apex. To associate this with our problem of interest, we normalize the length variables in Eqs. (9) and (10) with the cube root of the droplet volume $\sqrt[3]{V}$, which yields

$$\frac{d\phi}{d\hat{s}} = -\frac{\sin \phi}{\hat{x}} + \frac{2}{\hat{R}_0} - \tilde{F}_g \hat{z}, \tag{11a}$$

$$\frac{d\hat{x}}{d\hat{s}} = \cos \phi, \tag{11b}$$

$$\frac{d\hat{z}}{d\hat{s}} = \sin \phi, \tag{11c}$$

where $\tilde{F}_g = \frac{\rho g_{\text{eff}} V^{2/3}}{\gamma}$ is the same as the non-dimensional body force introduced in Eq. (5). After choosing the values of $\tilde{F}_g$ and $\hat{R}_0$, Eqs. (11a–c) can be numerically solved for $\phi(\hat{s})$, $\hat{z}(\hat{s})$, and $\hat{x}(\hat{s})$ using an ODE solver. The details on how Eqs. (11) were solved is provided in the Supplementary Information. Fig. 5(a) presents a typical parametric plot for $\hat{x}(\hat{s})$ and $\hat{z}(\hat{s})$, obtained from the solution for Eqs. (11a–c). The droplet shape then can be obtained by considering $\hat{R}_0$ to be a tunable parameter and solving Eqs. (11a–c) for $\{0 \leq \hat{s} \leq \hat{s}_{\max}\}$ with regard to the physical constraints $\hat{V} = \int_0^{\hat{s}_{\max}} \pi \hat{x}(\hat{s})^2 \sin \phi(\hat{s}) \, d\hat{s} = 1$ and $\phi(\hat{s}_{\max}) = \theta_R$. The first constraint is related to the fixed droplet volume, while the second is related to the receding contact angle at the liquid-solid-vapor interface. As shown in Fig. 5(b), the droplet shape reconstructed in this way is identical to that obtained from the LB simulations, which in turn are in correspondence with the experiments (Fig. 3).

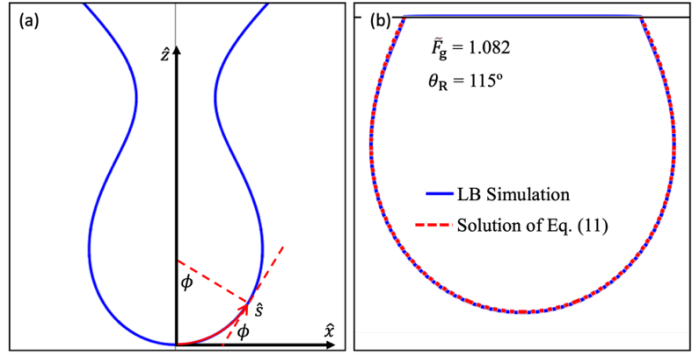

**Fig. 5. Determining the droplet shape from the YLE.** (a) Droplet liquid–vapor interface expressed in *x-z* coordinates as well as in path *s* and inclination angle $\phi$ coordinates. The interface shape is obtained from the parametric plot of $\hat{z}[\hat{s}]$ against $\hat{x}[\hat{s}]$ obtained from solving Eqs. (11). (b) A representative comparison of a pendant droplet contour reconstructed from LB simulations and from the solution for Eqs. (11) shows the agreement between the two methods.

Next, to determine $\tilde{F}_D$ using the YLE approach, we increase $\tilde{F}_g$ incrementally and determine the corresponding droplet shape by solving Eqs. (11) until the two constraints can no longer be satisfied simultaneously. For this specific $\tilde{F}_g$ value, the system cannot find a stable configuration, and the droplet becomes unstable and detaches. This is recorded as $\tilde{F}_D$. This strategy, however, is inefficient because $\hat{R}_0$ needs to be tuned for each $\tilde{F}_g$. A more efficient strategy for determining $\tilde{F}_D$ is by normalizing the length variables in Eqs. (9–10) with $R_0$ so that Eq. (11a) can be written as [45]

$$\frac{d\phi}{d\hat{s}} = -\frac{\sin\phi}{\hat{x}} + 2 - \beta\hat{z}, \tag{12}$$

where $\beta = \frac{\rho g_{\text{eff}} R_0^2}{\gamma}$, while Eqs. (11b) and (11c) are still in the same notation. In this case, the only tunable parameter is $\beta$ and the only physical constraint to determine the droplet shape is $\phi(\hat{s}_{\max}) = \theta_R$. By choosing $R_0 = 1$, $\beta$ is related to $\tilde{F}_g$ via $\tilde{F}_g = \beta \hat{V}^{2/3}$. Note: $\tilde{F}_D$ can be determined by finding the highest value of $\tilde{F}_g$ for different $\beta$, which indicates the maximum $\tilde{F}_g$ value for which the droplet is still stable. The results for determining $\tilde{F}_D$ using this strategy are displayed as black data points in Fig. 6, which accurately capture the experimental and simulation results. For practical reasons, the following model,

$$\tilde{F}_D = \frac{8.3}{1 + e^{0.033(\theta_R - 58.5)}}, \tag{13}$$

is fitted into the predicted data points. Eq. (13) then can be used to approximate $\tilde{F}_D$ of a given $\theta_R$ without the need to iteratively solve Eq. (12) numerically. Please note that, using the relation given in Eq. (6), calculating the detachment force in Newton is straightforward when the liquid properties ($\rho$, $\gamma$, and $V$) are known.

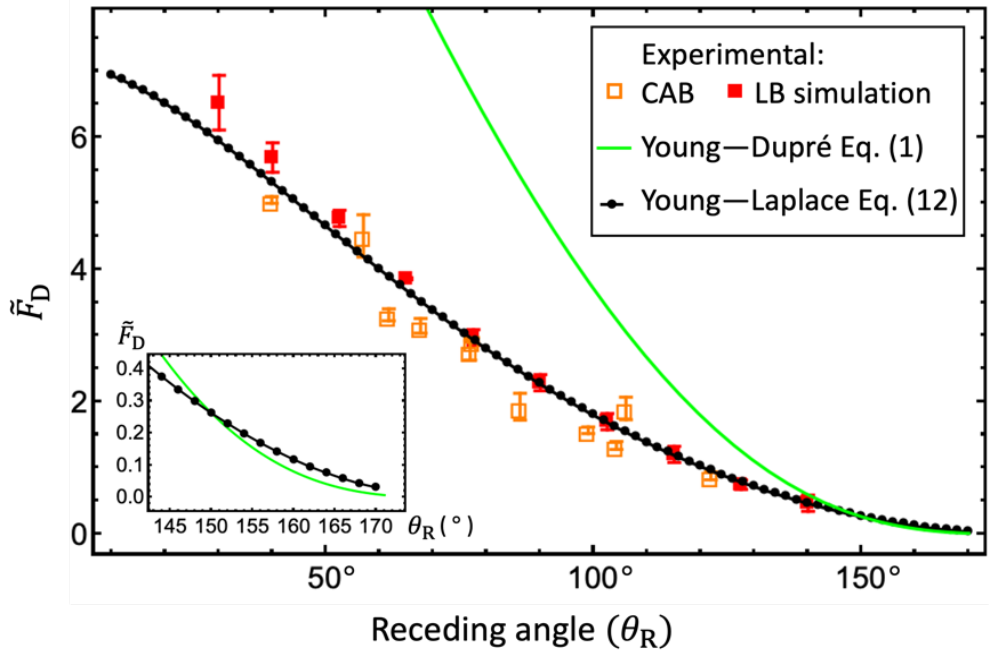

**Fig. 6. Theoretical prediction of the detachment force.** Comparison of the direct measurement of $\tilde{F}_D$ using CAB experiments (□) and LB simulations (■) and by solving Eq. (12), presented as (•). The data are also compared with the prediction based on the work of adhesion (Eq. 1), which is presented as (—).

**Generality of the YLE-based predictive framework:** We have demonstrated two approaches to predicting $\tilde{F}_D$ using the YLE with the volume constraint (Eq. 11) and without (Eq. 12). Although both approaches can accurately capture the results from CAB experiments and LB simulations (Fig. 6), the latter approach (Eq. 12) is more relevant to volume-induced detachment since the total droplet volume is not constrained. This begs the following fundamental question – *Could there be a critical volume or a critical interfacial tension that may also drive the droplet detachment akin to the gravity-induced detachment?*

To answer this, we designed LB simulations where the droplet volume was increased, or the interfacial tension was decreased, until the detachment occurs. Note: the latter case is relevant to displacing an oil droplet underwater by adding a surfactant. The results reveal an equivalence between droplet detachment realized by increasing gravity or increasing volume or decreasing the interfacial tension (Fig. 7). To our knowledge, this is the first time such an equivalence has been established.

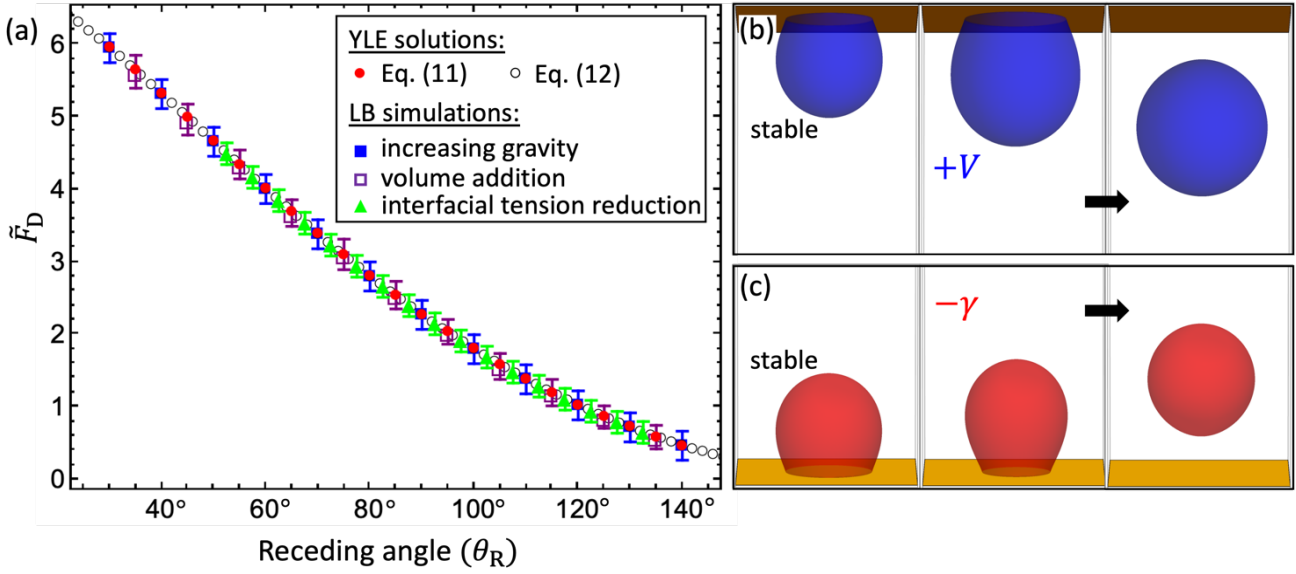

**Fig. 7. Testing the YLE predictions for three different simulation setups.** (a) A comparison of the detachment force, $\tilde{F}_D$, predictions from solving the YLE with and without volume constraint, respectively Eqs. (11) and (12). The theoretical predictions were also compared with LB simulations. Three simulation setups were used to determine $\tilde{F}_D$ including via increasing gravity (Fig. 3), (b) volume addition, and (c) interfacial tension reduction.

## III.    Discussion

This report unambiguously established that the droplet detachment force $\tilde{F}_D$ can be predicted accurately using the YLE. This equation describes stable pendant or sessile droplet shapes under an external force but, to the best of our knowledge, it has not previously been employed to recover $\tilde{F}_D$ [32]. Remarkably, when YLE was numerically solved in conjunction with constraints on the receding contact angle, the limit for the droplet stability could be determined; when the external force exceeded this critical value, detachment occurred. The YLE-predicted $\tilde{F}_D$ values were in excellent quantitative agreement with the $\tilde{F}_D$ values experimentally measured from the CAB over a wide range of chemical compositions on smooth and textured surfaces (Table 1) and those from LB simulations. The onset of this instability is an important event whose significance has not been fully recognized in the droplet detachment process, partly because most previous experimental studies have focused on hydrophobic and superhydrophobic surfaces.

To confirm that $\tilde{F}_D$ is not a function of $W_{SL}$ as previously believed, we compared the predictions from the YLE and Young–Dupre equation for $\tilde{F}_D$ over a wider range of $\theta_R$ (10°–170°) on smooth and flat surfaces. For liquid-repellent surfaces, when $\theta_R$ is 90°–150°, the predictions from the YLE closely matched the experimental and numerical observations, while the classical approach overpredicted $\tilde{F}_D$ (Fig. 6). Curiously, as $\theta_R$→150° and beyond, the two models appeared to converge qualitatively;

however, there were subtle quantitative differences that may be relevant in practical applications (inset of Fig. 6). It should also be noted that, whereas the YLE does not need to be modified to be applied to the analysis of droplet detachment from a microtextured surface, for the Young—Dupre approach, the fraction of solid in contact with liquid must be incorporated to account for air entrapment (Cassie—Baxter state). In Fig. 6, we limited the classic Young–Dupre predictions to smooth and flat cases only; because smooth surfaces with $\theta_R>110°$ are currently unavailable.

We also considered droplet detachment from wetting surfaces, where the complete detachment of the liquid droplet from the surface may not always be possible. As revealed by our experiments, the detachment process is characterized by necking followed by the break-up of the liquid column, leaving behind a residual droplet (SI Movie). The YLE predictions were in close agreement with the experiments, while the Young–Dupré equation overpredicted $\tilde{F}_D$ considerably. For extreme cases, as $\theta_R \rightarrow 0$, the YLE-predicted $\tilde{F}_D$ plateaus to a realistic value, while the values predicted by the Young–Dupre equation increase towards infinity because the droplet radius also increases toward infinity. This illustrates the limitations of the classic approach and the benefits of the YLE framework for droplet detachment. Notably, we also established an equivalence between detachment due to increasing gravity or increasing volume (for pendant droplets) or decreasing the interfacial tension (for sessile droplets).

Remarkably, a numerical solution for the YLE for a single data point required < 1 min on a typical laptop (1 CPU), whereas LB simulations required 2000–8000 CPU hours on a supercomputer for the same calculation. In addition to this, laboratory experiments take weeks and require significant human labor and research funding. This highlights the utility of the YLE for analyzing droplet detachment.

To conclude, this report advances the current understanding of droplet detachment physics, and the theoretical framework presented here will assist engineers in rationally designing surfaces with the appropriate chemical composition and microtexture, employing liquids to, for example, achieve complete droplet detachment, and/or tuning the relative volume of residual/detached droplets.

## IV. Methods

**Sample preparation:** Si wafers (p-doped, diameter of 10.2 cm, and thickness of 500 μm) and flat polystyrene were used as substrates. We grafted various silanes onto the silicon wafer following the method described in ref. [36]. Additionally, we microfabricated arrays of cylindrical pillars with a diameter of 20 μm, a height of 50 μm, and a pitch of 25 μm. We used photolithography and dry etching following the protocols reported in previous studies [46].

**Contact angle measurement:** Contact angles were measured using a goniometer (Krüss Drop Shape Analyzer DSA100). The advancing $\theta_A$ and receding $\theta_R$ contact angles were measured by adding 15 μl to a 2 μl droplet and removing it again at a rate of 0.2 μl/s. This process was repeated at a minimum of three different locations on each sample while recording images that were analyzed using *Advance* software (Krüss GmbH) to estimate the contact angles by fitting tangents at the solid–liquid–vapor interface.

**Centrifugal adhesion balance experiments:** The experimental droplet detachment force was measured using the Wet Scientific model CAB15G14. Each sample was cleaned by rinsing it with ethanol and water before each measurement. A liquid droplet was dispensed carefully in the middle of the sample using a micropipette. A body force was then increased incrementally until the droplet detaches from the sample.

**Lattice Boltzmann simulations:** The numerical investigation of the droplet detachment force was carried out using the free-energy LB method described in ref. [38]. The free-energy model is

$$\Psi = \int_V \left( \psi_b + \frac{\kappa}{2} (\nabla\phi)^2 \right) dV - \int_S h\phi_s \, dS, \tag{14}$$

$$\psi_b = \frac{c^2}{3} \rho \ln \rho + A \left( -\frac{1}{2}\phi^2 + \frac{1}{4}\phi^4 \right), \tag{15}$$

where $\rho$ is the fluid density, $\phi$ is the order parameter used as an identifier for the fluid phase, and $c = \Delta x/\Delta t$ represents the discretization of space and time. $A$, $\kappa$, and $h$ are tunable simulation parameters that set the surface tension $\gamma$ and contact angle $\theta$, respectively, as follows:

$$\gamma = \sqrt{8\kappa A/9}, \tag{16}$$

$$h = \sqrt{2\kappa A} \, \text{sign}\left(\frac{\pi}{2} - \theta\right) \sqrt{\cos\left(\frac{\cos^{-1}(\sin^2\theta)}{3}\right)\left\{1 - \cos\left(\frac{\cos^{-1}(\sin^2\theta)}{3}\right)\right\}}. \tag{17}$$

The hydrodynamics of the system are governed by the continuity, Navier–Stokes, and convection-diffusion equations:

$$\partial_t \rho + \vec{\nabla} \cdot (\rho \vec{v}) = 0, \qquad (18)$$

$$\partial_t (\rho \vec{v}) + \vec{\nabla} \cdot (\rho \vec{v} \otimes \vec{v}) = -\vec{\nabla} \cdot \mathbf{P} + \vec{\nabla} \cdot \left[ \eta \left( \vec{\nabla} \vec{v} + \vec{\nabla} \vec{v}^T \right) \right] + \rho \vec{g}, \qquad (19)$$

$$\partial_t \phi + \vec{\nabla} \cdot (\phi \vec{v}) = M \nabla^2 \mu. \qquad (20)$$

where $\vec{v}$, $\vec{g}$, and $\eta$ are respectively the fluid velocity, acceleration due to the body force, and viscosity. The free-energy model described in Eq. (14) enters the Navier–Stokes equation via the pressure tensor $\mathbf{P}$, whose form needs to satisfy the constraint $\partial_\beta P_{\alpha\beta} = \rho \partial_\alpha \left( \frac{\delta \Psi}{\delta \rho} \right) + \phi \partial_\alpha \left( \frac{\delta \Psi}{\delta \phi} \right)$. Eqs. (17–19) are then solved numerically using the LB algorithm [38, 47].

**CPU details:** For the LB simulations, two supercomputing nodes each with 32 CPUs with a 2.3 GHz processor speed and 128 GB of DDR4 memory running at 2300 MHz were employed. For the YLE approach, we utilized Mathematica software on a personal laptop equipped with a 2.3 GHz 8-Core Intel Core i9 processor and 16 GB 2667 MHz DDR4 RAM.

## Data availability

The datasets generated and/or analyzed during the current study are available from the corresponding author on reasonable request.

## Code availability

The free energy lattice Boltzmann code and Mathematica code used in the current study are available from the corresponding author on reasonable request.

## Acknowledgements

This study was supported by funding from King Abdullah University of Science and Technology (KAUST) under award number BAS/1/1070-01-01. The co-authors thank Mr. Edelberto Manalastas for building a new sample holder for the CAB, which simplified the experiments, Prof. Rafael Tadmor from Ben-Gurion University, Israel, for fruitful discussions, and Mr. Xavier Pita (KAUST) for the illustration presented in Figure 1A. MSS thanks Dr. Ciro Semprebon, Dr. Dan Daniel, and Dr. Meng Shi for fruitful discussions. This study used the computational resources of the Supercomputing Laboratory at King Abdullah University of Science & Technology (KAUST) in Thuwal, Saudi Arabia.

**Author Contributions:** H.M. designed and planed the experiments. M.S.S. performed LB simulations and data analysis. Y.X. prepared the samples and carried out CAB experiments. S.A. performed photolithography and provided the microtextured surfaces. M.S.S. and Y.X. developed the theoretical approach. M.S.S. and H.M. wrote the manuscript. All authors reviewed and approved the manuscript.

**Competing Interest Statement:** The authors declare no competing financial interests.